\documentclass[10pt,letterpaper]{article}
\usepackage{opex3}

\usepackage{graphicx}
\usepackage{cite}

\begin{document}

\title{Bragg reflection waveguide as a source of wavelength-multiplexed polarization-entangled photon pairs}

\author{Ji\v{r}\'{i} Svozil\'{i}k,$^{1,2}$ Martin Hendrych,$^{1}$ and Juan P. Torres$^{1,3}$}
\address{$^1$ICFO-Institut de Ciencies Fotoniques, UPC, Mediterranean
Technology Park, 08860 Castelldefels (Barcelona), Spain
\\ $^2$ Palack\'{y} University, RCPTM, Joint Laboratory of Optics,
17.listopadu 12, 771 46 Olomouc, Czech Republic
\\ $^3$Department of Signal Theory and Communications, Universitat
Polit{\`{e}}cnica de Catalunya, Castelldefels, 08860 Barcelona,
Spain}

\email{jiri.svozilik@icfo.es}

\begin{abstract}
We put forward a new highly efficient source of paired photons
entangled in polarization with an ultra-large bandwidth. The
photons are generated by means of a conveniently designed
spontaneous parametric down-conversion process in a semiconductor
type-II Bragg reflection waveguide. The proposed scheme aims at
being a key element of an integrated source of
polarization-entangled photon pairs highly suitable for its use in
a multi-user quantum-key-distribution system.
\end{abstract}


\section{Introduction}
Many quantum computing applications \cite{serafini2006,cirac1999}
and quantum communication protocols
\cite{ribordy2000,ekert1991,hillery1999}, such as quantum
teleportation \cite{bennet1993,humble2007}, are based on the
sharing of entangled two-photon states between two distant
receiving stations. The paired photons can be distributed through
free space transmission channels, or alternatively, through
single-mode fiber optical links. Indeed, fiber optical links are
the most practical way to share entanglement between a large
number of users.

The polarization degree of freedom is the most widely used
resource to generate entanglement between two distant parties.
Polarization-entangled photons can be generated by means of
spontaneous parametric down-conversion (SPDC), a nonlinear optical
process in which two lower-frequency photons (signal and idler)
are generated when a strong pump interacts with the atoms of a
nonlinear material. Subsequently to the generation, each photon
constituting a pair can propagate through a different single-mode
fiber, with the initial degree of entanglement between the photons
being preserved over long distances \cite{hubel2007}.

One application that is attracting a lot of interest due its
potential key role in the newly emerging quantum communication
networks is the multi-user quantum key distribution (QKD)
\cite{chapuran2009}. In order to implement a multi-user QKD
network, various frequency channels can expediently be employed
for transmitting individual entangled pairs. In this way, one can
re-route on demand specific channels between users located in
different sites of the optical network. Similar schemes,
considering the emission of photon pairs in different spectral and
spatial modes, have been presented in
\cite{migdall2002,shapiro2007} for an on-demand single-photon
source based on a single crystal.

To prepare polarization-entangled paired photons in many frequency
channels at the same time, one needs to engineer an SPDC process
with an ultra-broad spectrum. Usually type-I or type-0
configurations are preferred. With the type-II phase-matching, the
two down-converted photons have different polarizations and
consequently different group velocities, which reduces
dramatically their bandwidth. For instance, the FWHM bandwidth of
an SPDC process in a type-II periodically-poled (PP) KTP crystal
at $810$ nm is given by $\Delta \lambda (\rm{nm})=5.52/L({\rm
mm})$, where $L$ is the length of the crystal \cite{fedrizzi2007}.
For $L=1$ mm, the bandwidth is $\Delta \lambda \sim 5.5$ nm. On
the other hand, in a type-0 PPLN configuration with the same
crystal length $L=1$ mm, Lim et al. \cite{lim2008} achieved an
approximate tenfold increase of the bandwidth $\Delta \lambda \sim
50$ nm. Even though one can always reduce the length of the
nonlinear crystal in a type-II configuration to achieve an
increase of the bandwidth, this results in a reduction of the
spectral brightness of the source.

Alternatively to short bulk crystals, Bragg reflection waveguides
(BRWs) based on III-V ternary semiconductor alloys
(Al$_{x}$Ga$_{1-x}$As) offer the possibility to generate
polarization-entangled photons with an ultra-large bandwidth. The
most striking feature of the use of BRW as a photon source is the
capability of controlling the dispersive properties of all
interacting waves in the SPDC process, which in turn allows the
tailoring of the bandwidth of the down-converted photons: from
narrowband ($1-2$ nm) to ultra-broadband (hundreds of nm)
\cite{abolghasem2009,thyagarajan2008,Kang2012}, considering both
type-I and type-II configurations. Therefore, one can design a
type-II SPDC process in BRWs with a bandwidth typical for type-I
or type-0 processes.

The utilization of Bragg reflection waveguide (BRW) has further
advantages over other conventional SPDC sources due to the large
nonlinear coefficient of semiconductors ($d^{\rm GaAs}_{\rm eff}
\sim 119$ pm/V) \cite{shoji1997}, broad transparency window ($0.9$
- $17 \mu$m) and mature fabrication technologies that can be used
for integration of the source of entangled photons with a light
source and other optical elements. Additionally, a laser based on
BRWs has already been demonstrated \cite{biljani2009}, opening a
door for the integration in a single chip of the pump light source
together with the BRW, where the down-converted photons are
generated \cite{Horn2012}.

Al$_{x}$Ga$_{1-x}$As is an optically isotropic semiconductor,
precluding birefringent phase matching. However, the modal
phase-matching of the interacting waves (pump, signal and idler)
can be achieved by letting each wave propagate in a different type
of mode supported by the waveguide. For instance, the phase-matching
can be successfully achieved if the pump propagates in the waveguide in a Bragg
mode, whereas the signal and idler photons propagate in
total-internal-reflection (TIR) modes \cite{helmy2007}.

\begin{figure}[t]
\begin{center}
\includegraphics[width=0.9 \columnwidth]{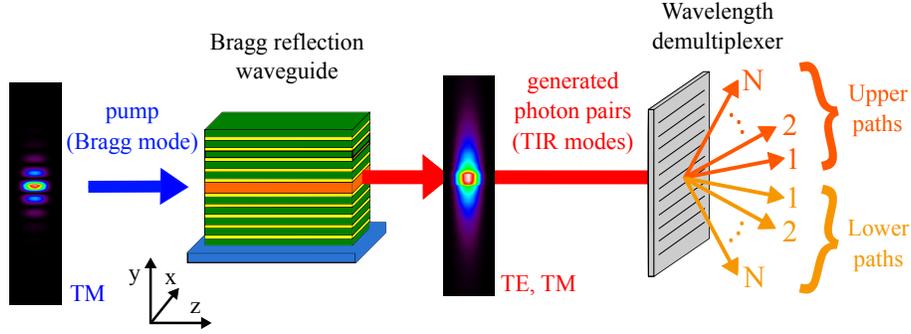}
\caption{General scheme for the generation of
polarization-entangled photon pairs in various frequency channels
by making use of the Bragg reflection waveguide. In this scheme, a
dichroic mirror or a grating can be used as the wavelength
demultiplexer.} \label{fig1}
\end{center}
\end{figure}

\section{Description of the quantum state of the down-converted photons}
In order to investigate the potential of the proposed design for
generating wavelength-multiplexed polarization-entangled photon
pairs over many frequency channels, let us examine the biphoton
generation in a collinear type-II phase-matching scheme in the
Bragg reflection waveguide (see Fig.~\ref{fig1}). A
continuous-wave TM-polarized pump beam with frequency $\omega_p$
illuminates the waveguide and mediates the generation of a pair of
photons with mutually orthogonal polarizations (signal: TE
polarization; idler: TM polarization). The frequencies of the
signal and idler photons are $\omega_s=\omega_0+\Omega$ and
$\omega_i=\omega_0-\Omega$, respectively, where $\omega_0$ is the
degenerate central angular frequency of both photons, and $\Omega$
is the angular frequency deviation from the central frequency. The
signal photon (TE) propagates as a TIR mode of the waveguide with
spatial shape $U_{s} (x,y,\omega_s)$ and propagation constant
$\beta_s(\omega_s)$. The idler photon (TM), also a TIR mode, has a
spatial shape $U_{i}(x,y,\omega_i)$ and propagation constant
$\beta_i(\omega_i)$. The pump beam is a Bragg mode of the
waveguide with spatial shape $U_{p}(x,y,\omega_p)$ and propagation
constant $\beta_p(\omega_i)$.

At the output face of the nonlinear waveguide, the quantum state
of the biphoton can be written as \cite{progress2011}
\begin{equation}
\label{state1}|\Psi_1\rangle=|\mathrm{vac}\rangle_{s} |\mathrm{vac}\rangle_{i} +
\sigma L F_p^{1/2}\int d\Omega \Phi\left(\Omega\right)
|\mathrm{TE},\omega_0+\Omega\rangle_{s} |\mathrm{TM},\omega_0-\Omega\rangle_{i},
\end{equation}
where the nonlinear coefficient $\sigma$ is defined
\begin{equation}
\label{sigma} \sigma=\left[ \frac{\hbar \omega_0^2  \omega_p
\left[ \chi^{(2)}\right]^2 \Gamma^2}{16 \pi \epsilon_0
c^3\,n_s(\omega_0) n_i(\omega_0) n_p(\omega_p)} \right]^{1/2}.
\end{equation}
$F_p$ is the flux rate of pump photons, $\Gamma=\int d{\bf
r}_{\perp} U_p({\bf r}_{\perp}) U_s^{*}({\bf r}_{\perp})
U_i^{*}({\bf r}_{\perp})$ is the overlap integral of the spatial
modes of all interacting waves in the transverse plane, and
$n_{p,s,i}$ are their refractive indices. The joint spectral
amplitude $\Phi(\Omega)$ has the form
\begin{equation}
\Phi (\Omega)={\rm sinc} \left[\Delta_k (\Omega)L/2\right] \exp
\left\{ i s_k (\Omega) L/2 \right\}.
\label{Eg3}
\end{equation}
The ket $|\mathrm{TE},\omega_0 + \Omega\rangle_s$
($|\mathrm{TM},\omega_0 - \Omega\rangle_i$) designates a signal
(idler) photon that propagates with polarization TE (TM) in a mode
of the waveguide with the spatial shape U$_s$ (U$_i$) and
frequency $\omega_0 + \Omega$ ($\omega_0 - \Omega$). The
phase-mismatch function reads $\Delta_k
(\Omega)=\beta_{p}-\beta_{s}(\Omega)-\beta_{i}(-\Omega)$, and $s_k
(\Omega)=\beta_{p}+\beta_{s}(\Omega)+\beta_{i}(-\Omega)$. The
function $|\Phi(\Omega)|^2$ is proportional to the probability of
detection of a photon with polarization TE and frequency
$\omega_0+\Omega$ in coincidence with a photon with TM
polarization and frequency $\omega_0-\Omega$.

After the waveguide, a wavelength demultiplexer is used to
separate all $n$ frequency channels into coupled fibers leading to
the users of the network. The bandwidth of each channel is $\Delta
\omega$ and their central frequencies are
$\omega_{U,L}^{n}=\omega_0 \pm n \Delta$, where $\Delta$ is the
inter-channel frequency spacing and the letter  U(L) indicates the
upper (lower) path (see Fig.~\ref{fig1}). After the demultiplexer,
the quantum state of the down-converted photons can be written as
\begin{eqnarray} \label{state2}
& & |\Psi_2\rangle=|\mathrm{vac}\rangle_{s} |\mathrm{vac}\rangle_{i} \nonumber \\
& & + 1/\sqrt{2}\,\,\sigma L F_p^{1/2}\,\int_{B_n} d\Omega \left\{
\Phi\left(\Omega\right)
|\mathrm{TE},\omega_0+\Omega\rangle_{U}|\mathrm{TM},\omega_0-\Omega\rangle_{L}
\right. \nonumber \\
& & \left. + \,\Phi\left(-\Omega\right)
|\mathrm{TM},\omega_0+\Omega\rangle_{U}|\mathrm{TE},\omega_0-\Omega\rangle_{L}
\right\},
\end{eqnarray}
where $\int_{B_n}$ designates the frequency bandwidth from
$\omega_{U,L}^n-\Delta \omega/2$ to $\omega_{U,L}^n+\Delta
\omega/2$ coupled into every single fiber. Since we are interested
in generating polarization-entangled paired photons coupled into
single-mode fibers, the signal and idler photons in the upper and
lower paths are projected into the fundamental mode ($U_0$) of the
fiber. The coupling efficiency between the signal and idler modes,
and the fundamental mode of the single-mode fiber are given by
$\Gamma_s=|\int dx dy U_{s}(x,y,\omega_s) U^*_0(x,y,\omega_0)|^2$
and $\Gamma_i=|\int dx dy U_{i}(x,y,\omega_i)
U^*_0(x,y,\omega_0)|^2$. They yield a value of $\Gamma_s=\Gamma_i
\approx 0.88$ in the whole bandwidth of interest, showing a
minimal frequency dependence. All the modes are normalized so that
$\int dx dy |U_{j}(x,y,\omega)|^2=1$ for $j=s,i,0$.

Neglecting the vacuum contribution in the final quantum state,
normalizing and tracing out the frequency degree of freedom, the
two-photon state can be represented by the following density
matrix, where we use the conventional ordering of rows and columns
as $\left\{
|\mathrm{TE}\rangle_U|\mathrm{TE}\rangle_L,|\mathrm{TE}\rangle_U|\mathrm{TM}\rangle_L,
|\mathrm{TM}\rangle_U|\mathrm{TE}\rangle_L,
|\mathrm{TM}\rangle_U|\mathrm{TM}\rangle_L \right\}$:
\begin{equation}
\label{state3}\rho_n=\left(%
\begin{array}{cccc}
  0   &    0           &    0      &    0 \\
  0   & \alpha_n       & \gamma_n  &    0 \\
  0   & \gamma_n^{*}   & \beta_n   &    0 \\
  0   &    0           &    0      &    0 \\
\end{array}%
\right),
\end{equation}
where
\begin{eqnarray}
& & \alpha_n=1/2\,\int_{B_n} d\Omega |\Phi(\Omega)|^2,
\nonumber \\
& & \beta_n=1/2\,\int_{B_n} d\Omega |\Phi(-\Omega)|^2,
\nonumber \\
& &  \gamma_n=1/2\, \int_{B_n} d\Omega \Phi\left(\Omega\right)
\Phi^{*}\left(-\Omega\right),
\end{eqnarray}
with $\rm{Tr[\rho_n]}=\alpha_n+\beta_n=1$.

\section{Numerical results}

\begin{table}[t]
 \caption{(a) Parameters of the structure: $t_{c}$ - core
thickness; $t_{1,2}$ - thicknesses of the alternating layers of
the Bragg reflector; $x_{c}$ - aluminium concentration in the
core; $x_{1,2}$ - aluminium concentrations in the reflector's
layers; $n_{c}$ - the refractive index in the core; $n_{1,2}$ -
refractive indices in the reflector's layers; ${\partial
\beta_{s(i)}}/{\partial \Omega}$ -the inverse group velocity of
the signal (idler) photon. The structure is optimized for the
collinear type-II SPDC. (b) Profile of the refractive index along
the \emph{y}-axis of the BRW.}
\begin{center}
(a)%
\begin{minipage}[c]{0.4\textwidth}%
\begin{center}

\begin{tabular}{c|c}
Parameter &  Value \\
\hline
 $t_{c}$ $\rm(nm)$  &   370 \\
$t_{1}$ $\rm(nm)$  &  127 \\
$t_{2}$ $\rm(nm)$  & 309 \\
$n_c(x_c=0.7)$ $ $  & 3.177 \\
$n_1(x_1=0.4)$ $ $  &  3.655 \\
$n_2(x_2=0.9)$ $ $  &  3.064   \\
Ridge width $\rm(nm)$ &  1770 \\
${\partial \beta_s}/{\partial \Omega}$ $\rm(ns/m)$  & 10.55 \\
${\partial \beta_i}/{\partial \Omega}$ $\rm(ns/m)$ &  10.56 \\
Waveguide length (mm) & 1 \\

\end{tabular}
\end{center}
\end{minipage}%
(b)%
\begin{minipage}[c]{0.4\textwidth}%
\begin{center}

\includegraphics[width=3.5cm,scale=1]{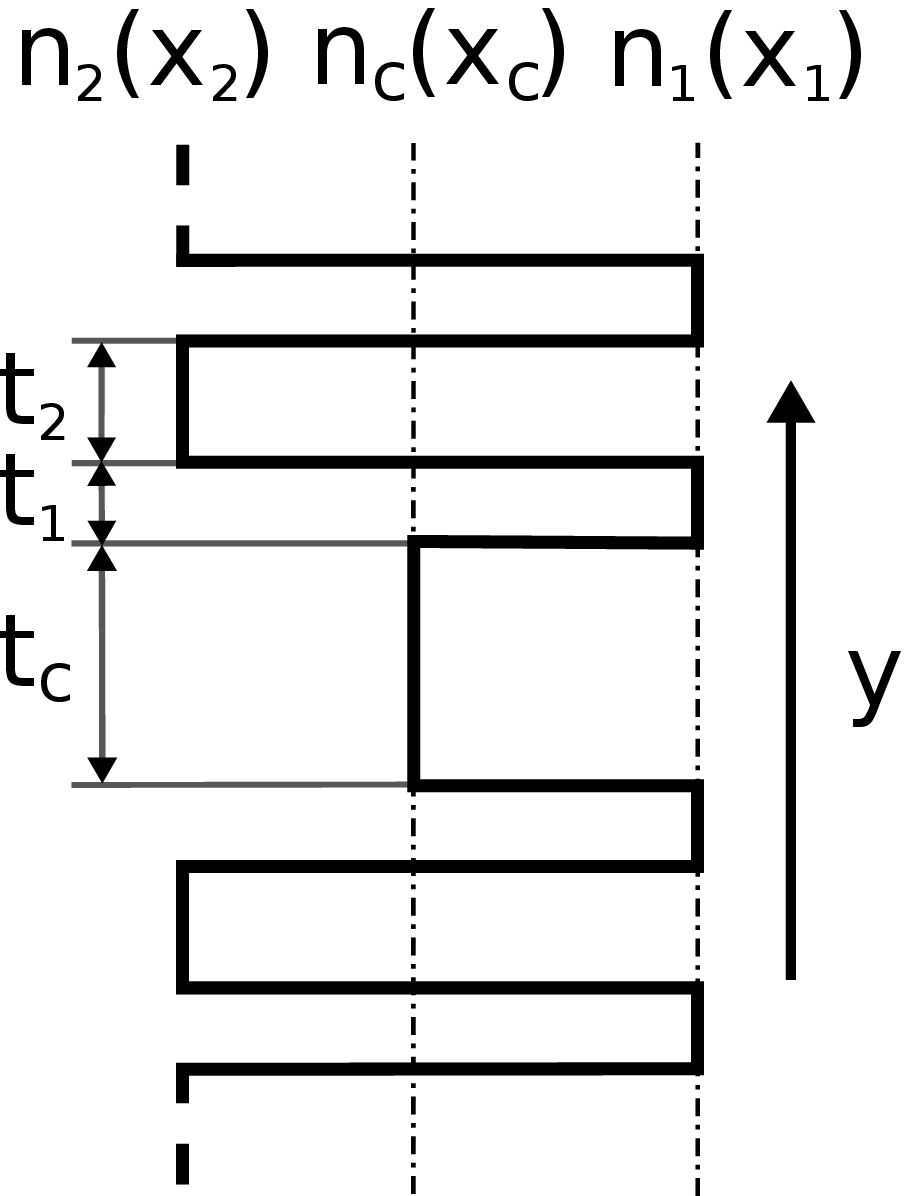}
\end{center}
\end{minipage}
\end{center}
 \label{fig_tab_1}

\end{table}

In the waveguide structure considered, the pump wavelength is
$775.1$ nm. The frequency spacing between channels is $\Delta=50$
GHz and the bandwidth of each channel is $50$ GHz, which
corresponds approximately to $0.4$ nm at $1550$ nm. The channel
width and the spacing between channels were chosen according to
the typical values used in commercial WDM systems. Channel $n=1$
corresponds to the wavelength $1549.6$ nm in the upper path and to
$1550.6$ nm in the lower path.

In order to reach a high number of frequency channels, the BRW
structure must be designed in such way, so as to permit the
generation of signal-idler pairs with an ultra-broad spectrum in
the type-II configuration. This is achieved when the group
velocities of the TE and TM modes are equal \cite{zhukovsky2012,torres2010},
i.e., $\vert\frac{\partial \beta_s}{\partial
\Omega}-\frac{\partial \beta_i}{\partial
\Omega}\vert\rightarrow0$. The modes of the structure and its
propagation constants are obtained as a numerical solution of  the
Maxwell equations inside the waveguide using the finite element
method \cite{Jianming2002FEM}. The waveguide design has been
optimized by a genetic algorithm according to the requirements.
The final BRW design  is composed of two Bragg reflectors, one
placed above and one below the core. Each reflector contains 8
bi-layers. The Sellmeier equations for the refractive indices of
the layers were taken from \cite{gehrsitz2000}. Table~1 summarizes
the main parameters of the structure.

\begin{figure}[t]
  \centering
  \includegraphics[scale=1.25]{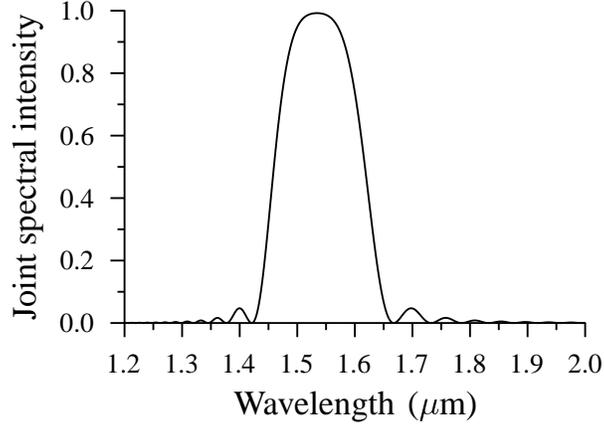}
  \caption{Joint spectral intensity $\sim|\Phi(\lambda)|^2$ of the biphoton
  generated in the Bragg reflection waveguide for type-II phase-matching ($\rm{TM \rightarrow TE+TM}$).}

  \label{fig2}
\end{figure}
Inspection of Eq.~(\ref{sigma}) shows that the effective
nonlinearity $\sigma$ of the waveguide SPDC process depends on the
effective area ($A_{\mathrm{eff}}=1/\Gamma^2$), which is related
to the spatial overlap of the pump, signal and idler modes. For
the structure considered, the effective area exhibits only minimal
frequency dependence in the bandwidth of interest and it is equal
to $A_{\mathrm{eff}}=35.3$~$\mathrm{\mu m^{2}}$. Despite the fact
that the large effective area will reduce the strength of the
interaction, the high nonlinear coefficient still results in an
efficiency that is a much higher than for other phase-matching
platforms in waveguides or in bulk media.  The total emission rate
\cite{Ling2008} can be expressed using Eq.~(\ref{state1}) as
$R=\sigma^2L^2\int d\Omega\vert\Phi\left(\Omega\right)\vert^2$.
For our BRW, the emission rate is
$R_{\mathrm{BRW}}\approx5.7\times10^7$~$\mathrm{photons/s/mW}$.
For comparison, for a typical PPLN waveguide (type-0) similar to
the one used in \cite{lim2008}, we obtain
$R_{\mathrm{PPLN}}\approx3.3\times10^7$~$\mathrm{photons/s/mW}$.
The intensity of the joint spectral amplitude, given by
Eq.~(\ref{Eg3}), is displayed in Fig.~\ref{fig2}. Even though we
are considering a type-II configuration, the width (FWHM) of the
spectrum is a staggering $160$ nm.

The degree of entanglement in each spectral channel can be
quantified by calculating the concurrence $C_n$ of the biphoton
\cite{hill1997,wooters1998}. The concurrence is equal to 0 for a
separable state and to 1 for a maximally entangled state. For the
density matrix of Eq.~(\ref{state3}) we obtain \cite{eberly2006}
\begin{equation}
C_n=2|\gamma_n|,
\end{equation}
so the degree of entanglement depends on the symmetry of the
spectral amplitude $\Phi\left(\Omega\right)$, i.e., if
$\Phi\left(\Omega\right)=\Phi\left(-\Omega\right)$ the concurrence
is maximum.

\begin{figure}[t]
  \centering
  \includegraphics[scale=1.25]{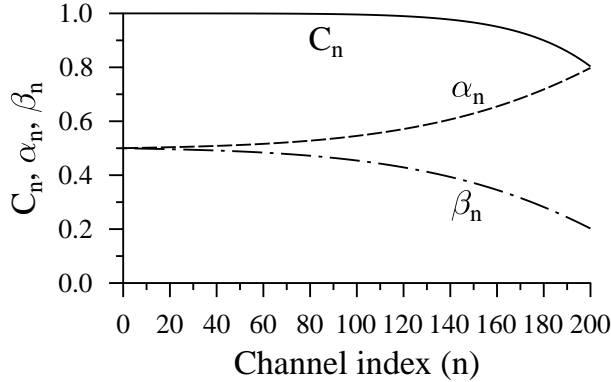}
  \caption{Coefficients $C_n$ (solid line), $\alpha_n$ (dashed line)
  and $\beta_n$(dotted-and-dashed line) as a function of the frequency channel.}
  \label{fig4}
\end{figure}

Figure~\ref{fig4} shows the values of $\alpha_n$, $\beta_n$ and
$C_n$ for the first 200 channels. $C_n > 0.9$ is reached for the
first 179 channels. The decrease (increase) of the parameters
$\beta_n$ ($\alpha_n$) reflects the fact that for frequency
channels with a large detuning from the central frequency, one of
the two polarization components of the polarization entangled
state, $|\mathrm{TE}\rangle_{1}|\mathrm{TM}\rangle_{2}$ or
$|\mathrm{TM}\rangle_{1}|\mathrm{TE}\rangle_{2}$,  shows a greater
amplitude probability. In this case, one of the two options
predominates. Therefore, if the goal is to generate a quantum
state of the form
$|\Psi_2\rangle=1/\sqrt{2}\,\left(\,|\mathrm{TE}\rangle_{1}|\mathrm{TM}\rangle_{2}
+|\mathrm{TM}\rangle_{1}|\mathrm{TE}\rangle_{2}\,\right)$ in a
specific frequency channel with $\alpha_n,\beta_n \ne 1/2$, one
can always modify the diagonal elements of the density matrix with
a linear transformation optical system, keeping unaltered the
degree of entanglement.

\begin{figure}[t]
  \centering
  \includegraphics[scale=1.25]{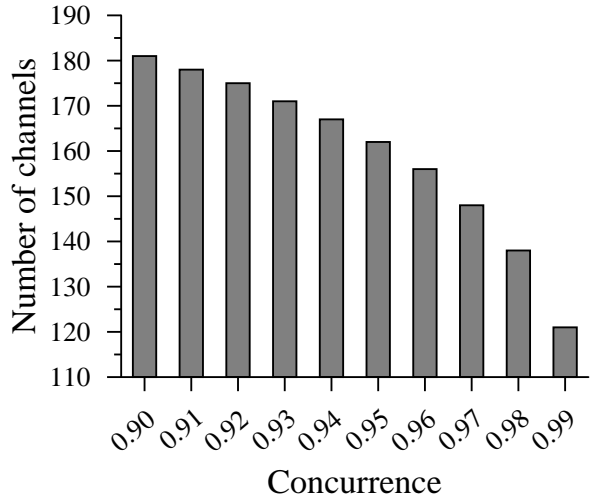}
  \caption{Number of channels available as a function of the minimum value of the concurrence required.}
  \label{fig5}
\end{figure}

The number of frequency channels available depends on the
concurrence required for the specific application. A good example
is a linear optical gate relying on the interference of photons on
a beam splitter \cite{kim2003,rohde2005}. This would be especially
important for the implementation of quantum teleportation, where
the fidelity of the protocol depends strongly on the spectral
indistinguishably between polarizations of the entangled state
\cite{humble2007}. In Fig.~\ref{fig5} we plot the number of
frequency channels available as a function of the minimum
concurrence required. For instance, if we select only frequency
channels with $C_n>0.95$, we have at our disposal $162$ channels,
while for $C_n>0.99$ this number is reduced to $121$ channels.

In the implementation of the system considered here in a {\em
real} fiber-optics network, the number of frequency channels
available can be limited by several factors. For instance, it can
be limited by the operational bandwidth of the demultiplexer (see
Fig.1). This device should be designed to operate with the same
broad spectral range of the photon pairs generated in the BRW
waveguide.

When using a large number of channels, inspection of Fig.
\ref{fig2} shows that channels far apart from the central
frequency will exhibit a lower brightness. In this case, spectral
shapers or appropriately designed filters, should be used to
flatten the emission spectrum, similarly to the case of broadband
gain-flattened Erbium doped fiber amplifiers (EDFA).
Notwithstanding, this might introduce some losses in the
generation process, especially when considering a large number of
channels, deteriorating the flux rate of the source.
Interestingly, a similar problem appears in the context of optical
coherence tomography (OCT), where large bandwidths are required to
increase the imaging resolution. In OCT, spectral shapers are used
to obtain an optimum (Gaussian-like) spectral shape \cite{oct}.

Finally, we should mention that the generation of
polarization-entangled photons with the large bandwidths
considered here require a precise control the group velocities of
the interacting waves, which in turn requires a precise control of
the waveguide parameters: refractive index and layer widths. The
effective number of available frequency channels in a specific
application is inevitably linked to the degree of control of the
fabrication process. Since both down-converted photons are
propagating as TIR modes, they are more resistant to fabrication
imperfections. For example, a change of about 10$\%$ in the
aluminium concentration in the core will reduce the spectral
bandwidth to $145$ nm. Notwithstanding, it has to be stressed that
the phase-matching condition for interacting waves is highly
sensitive to any fabrication imperfection, therefore any small
change of the structural parameters will lead to a shift of the
central (phase-matched) wavelength. 

\section{Conclusion}
In conclusion, we have presented a new type of highly efficient
waveguide source for generating polarization-entangled photon
pairs for its use in multi-frequency QKD networks. In spite of
being a type-II SPDC source, the achieved bandwidth is even larger
than the bandwidth usually obtained with type-I or type-0 sources.
The key enabling factor that allows us to achieve high efficiency
of the nonlinear process together with a bandwidth increase is the
fact that we can use a longer nonlinear material in a type-II
configuration, while at the same time keeping the broadband nature
of the SPDC process through the appropriate design of the Bragg
reflection waveguide structure.

Even though {\em conventional} sources based on the use of more
common nonlinear materials, such as KTP or LiNbO$_3$, might also
generate entangled pairs of photons with a large bandwidth, BRWs
based on AlGaAs compounds offer two main advantages: an enhanced
capability to tailor the general properties of the downconverted
photons, and the possibility of integration of different elements
(pump source, nonlinear waveguide and diverse optical elements) in
a chip platform based on an already mature technology, which could
pave the way for entanglement-based technologies in {\em
out-of-the-lab scenarios}.

\section*{Acknowledgments}
This work was supported by Project FIS2010-14831 and FET-Open
255914 (PHORBITECH).  J. S. thanks the project FI-DGR 2011 of the
Catalan Government. This work has also supported in part by
projects COST OC 09026, CZ.1.05/2.1.00/03.0058 of the Ministry of
Education, Youth and Sports of the Czech Republic and by projects
PrF-2011-009 and PrF-2012-003 of Palack\'{y} University.

\end{document}